# NEW MAGNETIC PHASE IN $Rb_3C_{60}$ FULLERENE IN NORMAL STATE OBSERVED IN TORQUE EXPERIMENTS.

J. Chigvinadze[1], V. Buntar[2] [*], S. Ashimov[1], T. Machaidze[1], and G. Donadze[1]
1. E. Andronikashvili Institute of Physics, 6 Tamarashvili str., 0177 Tbilisi, Republic of Georgia;
2. Columbia Int. College, 1003 Main Str. West, Hamilton, L8S 4P3 Ontario, Canada;
* *corresponding author (vbuntar@look.ca)*

Abstract

For the first time, magnetic properties of fullerides have been studied using a torque technique. $Rb_3C_{60}$ single crystal has been investigated in these experiments. It was shown that this method is sensitive to the structural phase transition from s.c. to f.c.c. structure. A rearrangement of magnetic system in this material was observed at temperature T ~ 200 - 250 K. Our results clearly show, that this effect is a crossover to an ordered magnetic phase related to interaction of magnetic dipoles and is related to the distorted T' site. The effect is discussed in terms of cooperative Jahn-Teller effect.

The appearance of superconductivity in alkali fullerides has led to extensive efforts in attempting to understand their electronic, structural, and dynamic properties and to elucidate the origin of their high $T_c$. In particular, the question whether or not such a large value of $T_c$ can be caused by coupling to phonon alone still to be answered and it strongly depends on the normal state properties. Despite the apparent simplicity of the structure of $A_3C_{60}$ fullerides, some important issues are not yet fully resolved. In this report we concentrate on discussion of two effects: the solid state transition from s.c. to f.c.c. structure and the possible transition/crossover connected to the distorted T' site.

a). *Solid state transition.* At high temperature, the solid of $A_3C_{60}$ form face-centered-cubic (f.c.c) phase. In this phase, $C_{60}$ molecules freely rotate with a reorientation time scale of the order of $10^{-11}$ seconds. With lowering temperature, the transition to s.c. structure occurs that comes from the fast reorientation of the $C_{60}$ molecules and an anisotropic uniaxial rotation of fullerene molecules takes place. In pure $C_{60}$ solid this transition occurs at T = 260 K and is well established. In alkali doped fullerenes, the precise structural transition temperature is not well established, but it is expected to be at higher T, since the presence of alkali metal molecules also prevents the free rotation of $C_{60}$'s. This transition has been observed for $K_3C_{60}$ by NMR experiments [1, 2], but exact transition temperature was hard to evaluate since, in these experiments, the peak related to this transition overlapped with a peak related to appearance of T' site and is very wide.

b). *Possible transition due to the structural distortion.* In the solid structure of fullerenes, there are two tetrahedral (T) and one octahedral (O) sites per $C_{60}$, where the sizes of T sites are smaller that that of O-site. In $A_3C_{60}$ structure per one fullerene molecule, one alkali ion occupies the octahedral site and two other occupy two tetrahedral sites. NMR studies show the expected T:O ratio of 2:1. At high temperature all T sites are equivalent. However, as it was found experimentally by $^{87}Rb$ NMR study [3] in $Rb_3C_{60}$ appears an additional resonance peak below 370 K. This peak is attributed to a second type of tetrahedral $Rb^+$ site, T', which is a modified T site. The "classical" ratio of O:T intensities 1:2 at T > 400 K is changed to O:T:T' = 35:55:10 at T = 200 K. As the result, the local symmetry is lower. Similar results have been reported for $K_3C_{60}$ [1, 4] and an additional peak has been observed on the temperature dependence of the spin lattice relaxation at temperatures around 200K. In [1] arguments have been presented that at this temperature the system overcomes a phase transition. Several models have been suggested, such as rotational orientation of $C_{60}$, i.e. freezing of $C_{60}$ molecular motion, Jahn-Teller distortion of $C_{60}^{3+}$, $Rb^+$ clustering, or carrier density modulation. In [5] was discussed that alkali-cation vacancies may provide an explanation for the T-T' splitting of the alkali-cation NMR lines. Unfortunately, there seems to be no consensus on the origin of this effect. However, since some reformation of the magnetic structure maybe expected, the torsion experiments maybe sensitive to this transition.

Direct measurements such as X-ray, transport measurements are particularly difficult to carry out on bulk samples, because they are very sensitive to air and have to be sealed in quartz or glass capsules. This inhibits the fabrication of suitable electrical contacts for transport measurements. As for X-ray, the sensitivity drops drastically down because of the capsules. Therefore, indirect magnetic methods are more suitable. The most common technique is SQUID magnetometer, which is very sensitive to existence of magnetic moments.

As an alternative to the SQUID magnetometer measurements, one may suggest the torsion technique that should be sensitive to the orientation of magnetic moments and, since each $C_{60}$ molecule may be considered as a magnetic dipole, to the reorientation of fullerene molecules in the normal state. This method is expected to be sensitive to the transition from simple cubic (s.c.) phase to the face-centered-cubic (f.c.c.) phase. We also expect that the torsion technique may provide some information about the transition, related to the appearance of T` site in $Rb_3C_{60}$.

In these experiments we apply the mechanic torque method in order to study the magnetic properties of alkali doped fullerenes. This method has been extensively used to investigate critical parameters of superconductors such as $T_c$, critical field $H_c$, energy dissipation in the mixed state [6, 7]. In the present research, we investigate magnetic phases in the normal state of the alkali doped fullerene $Rb_3C_{60}$.

A cylindrical (in the ideal case) sample is suspended on a thin elastic thread and performs axial-torsion oscillations in an external magnetic field, $\vec{H}$, which is perpendicular to the axis of the sample. The temperature dependence of the frequency, $\omega$, and of the dissipation of the oscillations, $\delta$, is measured at different magnitude of the magnetic field.

If there are no fixed (pinned) magnetic moments in the sample both the dissipation and the frequency of the oscillations do not depend on the external magnetic field. For example, when either i) external magnetic field does not penetrate the substance, which is the case of superconductor in the external field smaller than the lower critical field $H_{c1}$ for this material, or ii) inner magnetic moments are either zero or disoriented and not fixed.

Appearance of pinned magnetic dipoles produces nonzero magnetic moment $\vec{M}$ in the sample. The interaction between $\vec{M}$ and $\vec{H}$ makes a torque $\tau = MH\sin\alpha$, where α is the angle between $\vec{M}$ and $\vec{H}$. This additional moment τ affects the oscillating system and makes the dissipation and the frequency of oscillations dependent on the external magnetic field. The sensitivity of this method is very high, $10^{-17}$ W [7]. As it was shown by Galaiko [8], the interaction between pinned and unpinned vortices plays an important role in this process. A concentration of pinned and unpinned vortices changes with increasing of amplitude of oscillations, which leads to the amplitude dependence of the frequency ω and the dissipation δ. Pinning of vortices is responsible for the change of the frequency, while the interaction between pinned and unpinned vortices is responsible for the change of the dissipation. In this way, the suggested method allows to register appearance of a new magnetic phase and also to investigate its characteristics.

In our experiments we used a sample of $Rb_3C_{60}$ that was made from a single crystal of $C_{60}$ by doping it with Rb using the method of vapor phase doping (**). Details of the sample preparation and its characterization can be found in Refs. 9 and 10. The temperature dependence of the real (m') and the imaginary (m") parts of the ac susceptibility for this crystal are shown in Fig. 1 (***).

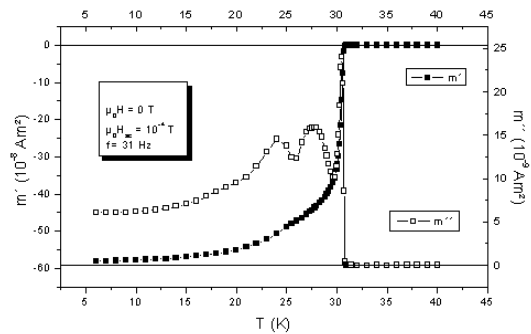

Fig. 1. Temperature dependence of the real m' (soild) and imaginary m'' (open) parts of the susceptibility obtained for $Rb_3C_{60}$ at an external magnetic field of $\mu_0H = 0.1$ mT.

The sharp drop of the magnetic moment shows the transition to the superconducting state at a temperature $T_c$ = 30.9 K. The peak at m" is very sharp and close to the critical temperature. The sharpness of the peak shows that the specimen is a bulk superconductor. At the same time, a complicated structure on m" below $T_c$ is observed. Several other peaks can be seen at lower temperatures. These peaks are attributed to dissipation due to weak links. Such granularity appears in the sample because of nonsuperconducting impurities (undoped $C_{60}$ or most probably $Rb_{x\neq3}C_{60}$) or as a block structure. Using the method suggested by Angadi et al [11] and later on developed for fullerene superconductors [12] we estimated the average size of the grains in this crystal to be R ≈ 350 μm. This very big sample (with dimensions approximately 3.3 x 2.7 x 1.3 $mm^3$) has obviously several big grains of the order of several tenth of a millimeter and, most probably, extensive granular regions. However, each grain may be considered as a single crystal of $Rb_3C_{60}$ with a good crystalline structure.

In this paper we address the dynamics of the magnetic system in $Rb_3C_{60}$ single crystal by studying the frequency and the dissipation of the oscillations of a sample in the temperature range from 5 K up to 300 K. In figure 2 temperature dependences of both the dissipation δ and the frequency ω of the oscillations are presented. The general behavior of the dissipation is very flat at the level of about $10^{-3}$ (see the insert of Fig. 3). The frequency though increases almost linearly up to T ~ 250 K. Above this temperature, ω tends to saturate. Since pinning of magnetic moments is responsible for the change of frequency (as discussed above), one may conclude that the number of fixed magnetic moments in the sample decreases with increasing temperature. As for the saturation at T > 250 K, it may be due to the superposition of several effects, such as the phase transition at T ~ 280

K and a crossover at T ~ 200 – 250 K.

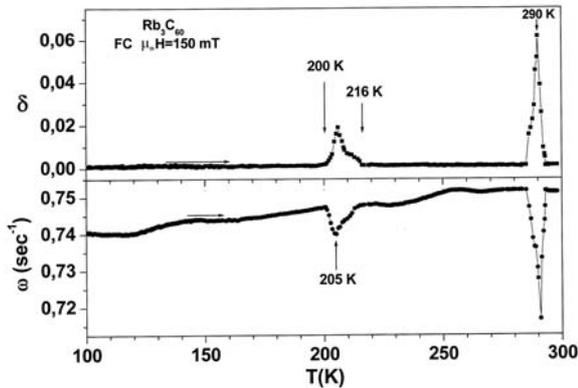

*Fig. 2. Temperature dependences of the dissipation δ (above) and the frequency ω (below) of the oscillations in $Rb_3C_{60}$ at external magnetic field $\mu_0 H = 150$ mT*

At around $T_1 \sim 280$ K a peak can be observed on both dependences. This peak may be associated with the first order phase transition from s.c. to f.c.c. structure. This transition is related to the freezing of libration modes and an orientation of $C_{60}$ molecules. If to consider each $C_{60}$ molecule as a diamagnetic dipole, then the orientation of molecules leads to the orientation of magnetic moments in the system and, therefore, strong response on δ(T) and ω(T). As it is expected, the transition temperature in $Rb_3C_{60}$ is higher than that in the pristine $C_{60}$ at T = 263 K due to the presence of Rb atoms in the interfullerene space.

There is a second peak at $T_2 \sim 200$-$250$ K. The amplitude of the peak of the dissipation (as shown in Fig. 3) is of several orders of magnitude, which is above of the sensitivity of our experimental device. Moreover, this peak is much larger that the one at s.c. – f.c.c. phase transition. We are not aware of any earlier experiments in which such strong magnetic response of fullerides material was observed and it was so obviously shown that there is a strong change in the magnetic system at this temperature.

The amplitude of the peak strongly depends on the prehistory of the measurements. In the first experiment (figure 3), we observe the giant dissipation. However, from measurement to measurement, this dissipation becomes smaller and, after 3-4 measurements, the peak is as big as it is shown in figure 2 and after few more measurements the peak practically disappears. Obviously that in order to return to the original level of the dissipation, the sample has to be annealed. A warming of the sample up to 300 K does not anneal the sample. Because of that, we applied a mechanical annealing. The sample was slowly rotated (from $0^0$ to $360^0$ and back) in external magnetic field of 150 mT at room temperature for more than 100 hours. The very first measurement after the mechanical annealing procedure again shows the giant dissipation.

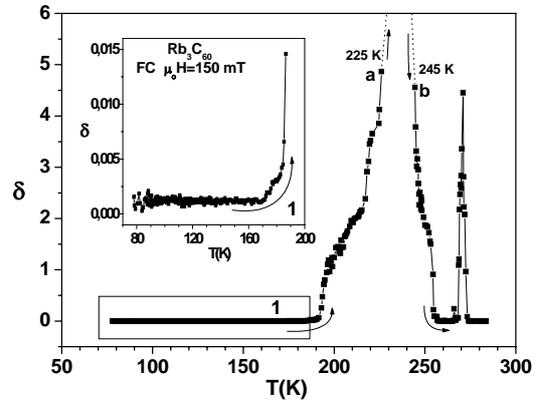

*Fig. 3. Temperature dependences of the dissipation δ of the oscillations in $Rb_3C_{60}$ at external magnetic field $\mu_0 H = 150$ mT taken in the first measurement after mechanical annealing*

In earlier NMR experiments, some anomalies on the temperature dependence of spin-lattice relaxation time have also been observed at this range of temperature in $K_3C_{60}$ [1, 2] and $Na_2CsC_{60}$ [13]. However, this anomaly was very weak and, as the authors reported, its amplitude was in the order of experimental error. In our experiments, the magnitude of the effect is huge. And, since it appears in our torque experiments which are sensitive to the presence of fixed/oriented magnetic moments, it means that at this temperature there is a reorganization of the magnetic structure in the material.

The fact that the effect cannot be annealed T ~ 300 K, which is 80 K above the position of the peak, is a strong argument that at $T_2 \sim 220$ K there is not a phase transition but a dynamical crossover. Also, since it cannot be annealed at T ~ 300 K, which is higher that the transition temperature to f.c.c. structure, it shows that the crossover is related to a process that starts at higher temperatures, most probably, to an appearance of the modified tetrahedral site T', which first shows up at T ~ 400 K [3]. The second fact, that the mechanical *rotation* anneals the sample indicates that the effect is due to the orientation of magnetic moments.

These magnetic moments cannot be the moments of $C_{60}$ molecules since most of the molecules are already oriented at $T_1 = 280$K and, moreover, magnetic effect is much stronger than that at the phase transition. Therefore, we may assume that the magnetic moments involved in the crossover are due to distortions of the $C_{60}$ molecules or the $A_3C_{60}$ lattice. As discussed in the previous paragraph, these distortions are, most probably, related to the appearance of T' site. As an explanation of the effect we suggest a scenario that is based on the Jahn-Teller effect, JTE.

In general, magnetism is mainly related to properties based on electron spins. The solid lattice, though important for some of effects, usually does not provide an important contribution to magnetism. However, there is a specific type of materials for which JTE plays very important role determining both structural and magnetic properties. In some compounds there are ions with an orbital degeneration (so-called JT ions) and, therefore, in symmetrical structural configuration the main state of magnetic ions has not only spin degeneration but, in addition, orbital degeneration. In these compounds, crystal structure is distorted and their magnetic structure is usually more complicated. Many classes of these materials have strong magnetic anisotropy and magnetostriction. High concentration of JT ions may lead to a cooperative JTE with interchange interaction between JT ions. The interchange interaction itself strongly depends on spins and leads to both orbital and spin ordering.

We suggest that T' site is a distorted structure around a JT ion. With temperature decreasing, number and concentration of T' sites and, therefore, JT ions is growing. At 400 K the ratio between O:T:T' intensities is 1:2:0. The corresponding ratio of intensities at 200 K is 35:55:10 [3]. At $T = T_2$ the density of T' sites reaches some critical value that leads to a cooperative JTE, which is a collective phenomena. From rough estimation, assuming that the ratio T:T' is 55:10, i.e. each sixth T site is the distorted T', there is one JT ion per three $C_{60}$ molecules. Therefore, we may estimate a distance between JT ions at $T_2$ around 2 – 3 nm. That is quite reasonable distance to create a collective effect. Also, from the figure 2, if to compare the amplitude of the peak relative to the "original" value of ($\delta$) $\omega$ just before (or after) the transition, one can find that $\Delta\delta$ is of the order of 40 times, while $\Delta\omega$ is about 4%. From here one can conclude that during the crossover, there is almost no change in the number of pinned magnetic moments but instead appears an interaction between the magnetic moments.

In addition, investigations of cooperative JTE in different materials (for a review see Ref. 14) show that even at temperature that is higher then the temperature of the cooperative JT adjustment, there are local distortions close to JT centers and they are chaotically distributed in the lattice. This scenario is very similar to what one can see in $A_3C_{60}$.

In summary, in this work we investigate magnetic phases in the normal state in $Rb_3C_{60}$. It is found that there are two rearrangements in the magnetic system at T ~ 280 K and 200 K < T < 250 K. The first one, at higher temperature, is the phase transition due to the freezing of a free rotation of $C_{60}$ molecules. It is shown that at lower temperature there is a crossover due to orientation of magnetic moments that are not related to the rotation of fullerenes but to distortions of the lattice. This effect is discussed in terms of cooperative Jahn-Teller effect.